\documentclass[aps,pre,twocolumn]{revtex4}
\usepackage{graphicx}

\begin{document}

\newcommand{\PE}{{\mathcal V}}
\newcommand{\fexp}{\alpha}
\newcommand{\Kzexp}{c}
\newcommand{\Pzexp}{e}
\newcommand{\Vzexp}{g}
\newcommand{\Sm}{\scriptstyle}
\newcommand{\polyexp}{\gamma}
\newcommand{\polyexptwo}{\nu}

\title{A mean field description of jamming in non--cohesive frictionless particulate systems}

\author{D. A. Head}
\affiliation{Division of Physics and Astronomy, Vrije Universiteit, Amsterdam, The Netherlands.\\
Department of Applied Physics, School of Engineering, University of Tokyo, Japan.}

\date{\today}

\begin{abstract}
A theory for kinetic arrest in isotropic systems of repulsive, radially--interacting particles is presented that predicts exponents for the scaling of various macroscopic quantities near the rigidity transition that are in agreement with simulations, including the non--trivial shear exponent. Both statics and dynamics are treated in a simplified, one--particle level description, and coupled {\em via} the assumption that kinetic arrest occurs on the boundary between mechanically stable and unstable regions of the static parameter diagram. This suggests the arrested states observed in simulations are at (or near) an elastic buckling transition. Some additional numerical evidence to confirm the scaling of microscopic quantities is also provided.
\end{abstract}


\maketitle

\section{Introduction}
\label{s:intro}

Determining when and how a self--assembled particulate system is capable of supporting macroscopic loads appears to be becoming established as a core paradigm in non--equilibrium condensed matter. For many dissipative athermal systems, such as granular media~\cite{Cates:98,Geng:03,Moukarzel:04,Kasahara:04a,Kasahara:04b,Barker:93,Head:00}, foams~\cite{Rafi:DryFoam,Durian:Foam,Bolton:90} or emulsions~\cite{Mason:95,Zhang:05}, the removal of a heat bath or other energy source results in a monotonic decrease of particle motion until a state of kinetic arrest is reached. The morphology, and hence mechanical response, of this static phase is determined by the dynamics of its preparation~\cite{Cates:98}, demonstrating the central importance of the initial dynamic phase and highlighting the non--equilibrium nature of the problem. It is this aspect that delineates self--assembled systems to those with a predefined geometry, such as the tensegrity structures of great importance to engineering~\cite{Calladine:78,Connelly:96,Guest:03}, or the disordered discrete and continuous models of rigidity percolation~\cite{Sahimi:Book,Jacobs:95,Astrom:00,LatvaKokko:01,Head:03,Bergman:84,Feng:84,Arbabi:88,Tang:88,Zhou:03}.

A generic feature of these distinct but related problems is the existence of a {\em rigidity} transition between states that have non--zero elastic moduli, and those that do not~\cite{footnote,Donev:04,OHern:04,OHern:03,Aharonov:99,Makse:EMAFail,Makse:05,Donev_condmat}. For bond diluted lattices, the rigidity transition corresponds to a critical bond dilution, or equivalently a critical mean coordination number~$z_{\rm c}$, that is well approximated by the Maxwell constraint counting method~\cite{Calladine:78}. This transition appears to bear some of the hallmarks of a continuous phase transition in equilibrium systems, including a diverging length scale, although other properties such as universality remain unproven~\cite{Head:03}. For particulate systems, the picture is less clear. Central force models, such as frictionless elastic spheres in the limit of small deformation, appear to reach a similar transition~\cite{Kasahara:04b,OHern:03}, also with a diverging length scale~\cite{Silbert:05}, in the limit of infinite interaction stiffness under controlled pressure. If the volume is controlled, a critical volume fraction must be fine tuned. However, the introduction of friction, at least, appears to overshoot the critical $z_{\rm c}$ by at least a few percent when in gravity~\cite{Kasahara:04a,Kasahara:04b}, the exact amount apparently depending on the damping.

\begin{figure}
\centering
\includegraphics[width=\columnwidth]{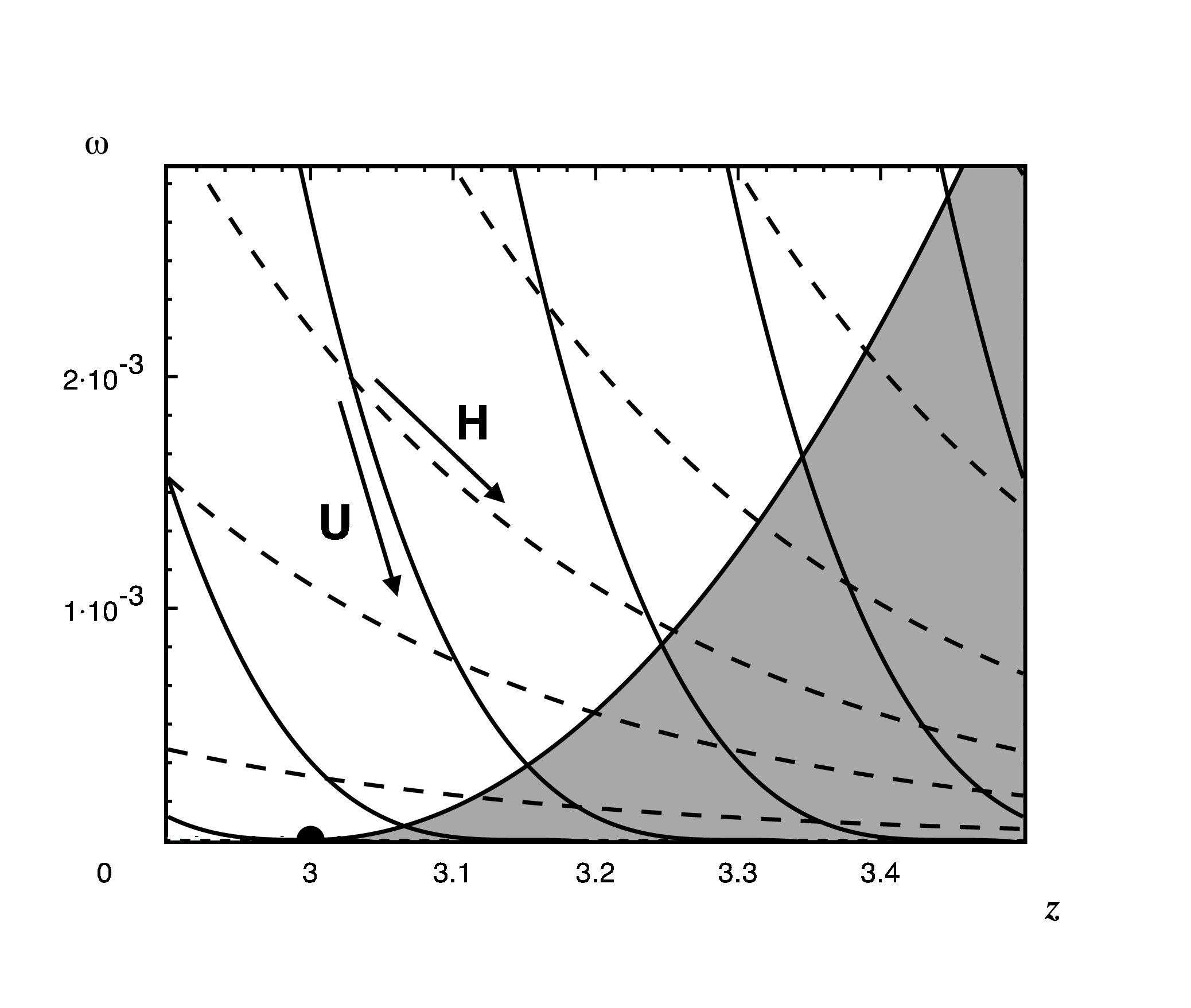}
\caption{Combined dynamic flow and static stability diagrams in $(z,\omega)$ space, with $z$ the mean coordination number and $\omega>0$ a quantity proportional to the mean particle overlap. Solid (dashed) lines correspond to lines of constant volume $V$ (pressure $P$), with the direction of minimising internal energy $U$ (enthalpy $H$) given by the arrows. Lines to the lower left corner correspond to smaller pressure or larger volume. The system becomes kinetically arrested at the boundary of the (shaded) stable region. The transition point at $(z_{\rm c},\omega)$ is shown as a solid disc (see text for discussion of $z_{\rm c}$).}
\label{f:Min_PV}
\end{figure}

Even when the transition can be approached arbitrarily closely, there remain crucial differences between disordered lattices and particulate systems. In particulate systems, the morphology of the solid state depends on the dynamics of the immediate precursor to arrest, and may therefore include non--trivial, correlated structures. Most studies of lattice models assume morhpologies with no structure beyond the one--bond level. Furthermore particles may become arrested in a state with a finite pressure $P>0$, quite unlike the stressless configurations typically adopted in lattice models.  Attempts have been made to circumvent some of the deficiencies of lattice models by postulating initial configuration generating algorithms that are hoped to mimic the dynamic self--assembly of particulate or atomic systems. Indeed one such scheme, known as bootstrap percolation, has already been applied to this problem~\cite{Schwarz:05}, and by infinite dimensional calculations claim to determine the exponents observed in simulations of central force systems.

The purpose of this paper is to describe an analytical scheme that, when applied to isotropic systems of particles interacting {\em via} central forces, derives the exponents relating pressure, shear modulus and volume fraction to $z-z_{\rm c}$ observed in simulations. There is no mapping to a known model, or assumption of any statistical mechanical analogy, as recently suggested by the elegant theory of  Henkes {\em et al.}~\cite{Henkes:05}. Instead, the properties of the static system are approximated according to a minimum assumption philosophy that proposes a series of intuitive approximations to locally close the equations. For the dynamics, a one--contact level closure is employed, whereas for the statics the approximation takes place somewhere between the one--contact and one--coordination shell level, in that it considers all contacts acting on a particle but ignores correlations between them. The static aspect of this scheme  has already been applied to mixed tensile--compressive systems~\cite{Head:04}.

The central result of this paper is schematically represented in Fig.~\ref{f:Min_PV}, in which the statics and dynamics of the theory have been overlayed into a single plot. The axes are $z$ and $\omega$, where $\omega$ is proportional to the mean particle overlap, so higher $\omega$ corresponds to increasingly compressive contacts. The rigidity transition lies at $z=z_{\rm c}$, $\omega=0$, which corresponds to the same unstressed transition observed in {\em e.g.} lattice models. The anomalous low $z_{\rm c}=3$ (which should be closer to $2d=6$ in this $d=3$ dimension example) is explained later, but is essentially due to the reduction of degrees of freedom inherent in the approximation scheme. The shaded region corresponds to systems that are mechanically stable in that they generate a positive restoring force in response to a localised, linear perturbation. The upper boundary of this stable region, which represents some form of buckling, is locally quadratic, and it is this shape that ultimately determines the exponents.

This static picture is augmented by a global description of the system dynamics from excited to arrested states. Again approximations are required, in this instance the kinetic energy is ignored and the system is assumed to evolve towards state with a lower internal energy (in the case of fixed volume) or enthalpy (in the case of fixed pressure), not unlike thermal systems~\cite{Weiner:Book} but without temperature/entropy terms. Some choice is required in defining volume in terms of $z$ and~$\omega$, but for rapid quenches from highly excited states, a single--particle volume function should be valid, and will robustly give the direction given in the figure. This drives the system towards the stable region. Kinetic arrest is assumed to take place when the boundary of the stable region has been reached; that is, the configurations observed in simulations lie on a line of buckling transitions. By suitably controlling pressure or volume, it is possible to bring the system to the same rigidity transition as in stressless systems, but along a different line to the $\omega\equiv0$ systems considered in disordered lattices.


\section{Statics: The mean mode approximation}
\label{s:mma}

A variety of approximate analytical treatments for predicting the mechanical response of athermal materials have been devised; just two will be mentioned here by way of comparison. Perhaps the most straightforward is to assume the induced deformation field is affine, so the interparticle displacements are just scaled--down versions of the macroscopic field. This has been applied to frictional granular media~\cite{Walton:87}, but is incapable of predicting a rigidity transition at finite density or volume fraction. Even side--stepping this deficiency by directly comparing elastic moduli to pressure does not resolve its failings~\cite{Makse:EMAFail}, indicating that modes near the transition are inherently non--affine. Enhanced versions have been proposed in which additional degrees of freedom are introduced and determined by variational principles~\cite{Kruyt:98,Trentadue:01}, but only at the cost of significant additional complexity and still not suitable for studying the transition.

Another approach, extensively applied to unstressed systems of predefined morphology, is known as EMA or the {\em effective medium approximation}  (although this term is sometimes also applied to the affine approximation~\cite{Walton:87}). Here the disordered system is replaced by a homogenous analogue with effective interaction parameters. The procedure can be crudely summarised as follows: a single disordered element is inserted into the homogenous bulk and the response to this isolated defect determined using the Green's function. The effective parameters are determined by demanding that the mean response averaged over all possible states of the inserted element is zero. Although this has been successfully applied to a range of continuous and lattice systems~\cite{Sahimi:Book,Feng:85,Schwartz:85,Astrom:00}, it has a fatal shortcoming when applied to particulate systems: the calculation cannot proceed without first identifying an analogous homogeneous material with a known Green's function. It is difficult to postulate such an analogue for finite--size particles, except in certain special cases such as quasi--ordered systems~\cite{Velicky:02}. Furthermore we have already argued that prestress is likely to be important, and even for lattice spring networks the Green's function with prestresses is not known exactly~\cite{Tang:88}. For granular media the problem is particularly acute, as the Green's response has been the subject of substantial debate in recent years (see {\em e.g.}~\cite{Cates:98,Geng:03,Moukarzel:04} and refs therein). The approximation scheme detailed below circumvents these issues, as we describe below.

For this first exposition of the theory, the non--damping part of the interparticle interaction is assumed to be a purely radial pair potential, generating central forces acting along lines connecting particle centres. This choice is made to reduce the parameter space to a manageable size and to allow a systematic and lucid unfolding of the resulting phase structure of the system. Nonetheless this simplification should hold approximately true for emulsions or wet foams near the transition, when all particles are only slightly deformed, although it will clearly not apply to strongly deformed particles (or dry foams~\cite{Rafi:DryFoam}). For real granular media there will also arise transverse surface forces mediated by non--zero friction at the bead--bead interface; purely central forces correspond to frictionless spherical beads, which only truly exist in numerical simulations.

\subsection{Determining the mechanical stability}

Given central force interactions, there is no need to track particle orientations and the static system is fully specified by the $d$--dimensional position vectors ${\bf x}^{\beta}$ of all particles $\beta$. The force on $\beta$ due to $\gamma$ is denoted by~${\bf f}^{\gamma\beta}$,

\begin{equation}
{\bf f}^{\gamma\beta}
=
f(r^{\gamma\beta}){\hat {\bf n}}^{\gamma\beta}
\end{equation}

\noindent{}where $r^{\gamma\beta}=\mid{\bf x}^{\beta}-{\bf x}^{\gamma}\mid$ is the distance between particle centres and ${\hat {\bf n}}^{\gamma\beta}=({\bf x}^{\beta}-{\bf x}^{\gamma})/r^{\gamma\beta}$ is the unit vector from $\gamma$ to $\beta$. In this context the scalar central force $f(r)$ is usually taken to be of the form

\begin{equation}
f(r)
=
\left\{
\begin{array}{lll}
\mu\left(1-\frac{r}{r_{0}}\right)^{\fexp}
& : & r<r_{0} \\
0 & : & {\rm otherwise}
\end{array}
\right.
\label{e:f_r}
\end{equation}

\noindent{}with $\fexp=1$ (truncated Hookean) or $\fexp=3/2$ (Hertzian), and $r_{0}$ is the sum of the two particle radii. The prefactor $\mu>0$ is typically treated as a particle--independent parameter, although strictly speaking it is a function of the radii of curvature at the point of contact for Hertzian interactions~\cite{LandauLifshitz,Makse:05}. Note that with this sign convention, positive $f$ corresponds to compressive forces and negative $f$ to tensile ones.

Suppose we are given an initial configuration $\{{\bf x}^{\beta}\}$ that is static, {\em i.e.} the vector sum of all contact forces on each particle vanishes. To determine its linear stability, apply an arbitrarily small external force $\delta{\bf f}^{\rm ext}$ onto the particle lying nearest some arbitrary point in space --- call this particle~$\alpha$ (not to be confused with the force law exponent). If the system is stable, it will move to a nearby static configuration in which all particles $\beta$ have been displaced to ${\bf x}^{\beta}+\delta{\bf x}^{\beta}$ with $|\delta{\bf x}|\ll r_{0}$. Force balance must again be obeyed, {\em i.e.} the changes in contact forces on $\beta$ sum to zero for $\beta\neq\alpha$, and to $-\delta{\bf f}^{\rm ext}$ for particle~$\alpha$.

The response $\delta{\bf x}^{\alpha}$ for a particular configuration $\{{\bf x}^{\beta}\}$, even if tractable, would be of no practical interest and we must instead ensemble average, keeping fixed a set of parameters that dominate the mechanical response of the system. A wealth of data has shown the mean coordination number $z$ to be a crucial factor in determining stability, and Alexander~\cite{AlexanderRev} has highlighted the importance of prestresses, so we also assume that both $z$ and some measure of the initial contact force distribution is kept fixed.  Once the ensemble (denoted by the angled brackets $\langle\ldots\rangle$ below) has been suitably defined, the requirement of force balance on the perturbed bead $\alpha$ can be written

\begin{equation}
\delta {\bf f}^{\rm ext}
-
\left\langle
\sum_{\beta\sim\alpha}
\delta{\bf f}^{\alpha\beta}
\right\rangle
=
0
\label{e:force_balance}
\end{equation}

\noindent{}where the sum is over all $\beta$ interacting with~$\alpha$. The change in contact force $\delta{\bf f}^{\alpha\beta}$ can be related to the particle displacements $\delta{\bf x}^{\alpha}$ and $\delta{\bf x}^{\beta}$ by

\begin{equation}
\delta f^{\alpha\beta}_{i}
=
A^{\alpha\beta}_{ij}
\left(
\delta x^{\beta}_{j}
-
\delta x^{\alpha}_{j}
\right)
\label{e:delta_f_ab}
\end{equation}

\noindent{}with summation over Roman indices only. The $d\times d$ matrix $A^{\alpha\beta}$ is defined by~\cite{Tanguy:02}

\begin{equation}
A_{ij}=
\frac{f(r)}{r}
(\delta_{ij}-\hat{n}_{i}\hat{n}_{j})
+f^{\prime}(r)
\hat{n}_{i}\hat{n}_{j}
\label{e:A_ij}
\end{equation}

\noindent{}assuming $f(r)$ is continuous with a finite first derivative $f^{\prime}(r)\neq0$ over all $r$ of interest. Here and below the suffices $\alpha$, $\beta$ are dropped whenever the meaning is clear.

\subsection{Derivation of the MMA}


So far this is exact but intractable. Now we approximate. Consider that, for an isotropic system, the perturbed bead must move parallel to the external force after averaging, $\langle\delta{\bf x}^{\alpha}\rangle=\lambda\delta{\bf f}^{\rm ext}$ with an unknown compliance~$\lambda$. The philosophy of the MMA is to impose this form {\em before} averaging, {\em i.e.} inside the brackets in~(\ref{e:force_balance}). In this way the dependency of $\delta{\bf x}^{\alpha}$ on the entire initial configuration $\{ {\bf x}^{\gamma}\}$ is subsumed into the single scalar parameter~$\lambda$. This is clearly a significant saving in terms of complexity, although it disallows the transverse motion of the particle and hence reduces the degrees of freedom; the consequence of this on the location of the rigidity transition will be discussed later.

The logical continuation of this approach is to similarly replace the $\delta{\bf x}^{\beta}$ by $\langle\delta{\bf x}^{\beta}\rangle$; however, this averaged form cannot be determined by symmetry considerations alone. Instead we assume here that the change in the contact force with $\alpha$ can be treated as an external force on~$\beta$, so that $\delta{\bf x}^{\beta}=\lambda\delta{\bf f}^{\alpha\beta}$ with the same $\lambda$ as before. Intuitively, this corresponds to the statement that the displacement of $\beta$ is dominated by the change in contact force with $\alpha$, which, for a monotonically decaying force field extending outwards from~$\alpha$, should at least not be embarrassingly wrong. These two approximations taken together allows each contact force $\delta{\bf f}^{\alpha\beta}$ to be uniquely determined from $\delta{\bf f}^{\rm ext}$, as found by inserting $\delta{\bf x}^{\alpha}=\lambda\delta{\bf f}^{\rm ext}$ and $\delta{\bf x}^{\beta}=\lambda\delta{\bf f}^{\alpha\beta}$ into (\ref{e:delta_f_ab}) and (\ref{e:A_ij}) and inverting,

\begin{eqnarray}
\delta f^{\alpha\beta}_{i}
&=&
S^{\alpha\beta}_{ij}
\delta f^{\rm ext}_{j},
\nonumber\\
S^{\alpha\beta}_{ij}
&=&
\left[
1+(\lambda|f^{\prime}(r^{\alpha\beta})|)^{-1}
\right]^{-1}
{\hat n}^{\alpha\beta}_{i}{\hat n}^{\alpha\beta}_{j}
\nonumber\\
&&+
\left[
1-\left(
\frac{\lambda f(r^{\alpha\beta})}
{r^{\alpha\beta}}
\right)^{-1}
\right]^{-1}
\left(\delta_{ij}-{\hat n}^{\alpha\beta}_{i}{\hat n}^{\alpha\beta}_{j}\right)
\nonumber\\
\label{e:Sij}
\end{eqnarray}

\noindent{}Thus each $\delta{\bf f}^{\alpha\beta}$ is independent of the others. Note that the unphysical singularity at $\lambda f/r=1$ is avoided by the stability equation below.

It is apparent from (\ref{e:Sij}) that the MMA has reduced the global problem to a local one, in which the response of each contact $\delta{\bf f}^{\alpha\beta}$ depends only on the interparticle separation $r^{\alpha\beta}$ (through which $f(r)$ and $f^{\prime}(r)$ are found), the unit vector $\hat{\bf n}^{\alpha\beta}$, and the coordination number for this bead $z^{\alpha}$, implicit in the summation~(\ref{e:force_balance}). Before the averaging can be completed, it is necessary to specify how these quantities vary. Here we will deliberately take simple forms to facilitate transparent interpretation of the results. Firstly, $z^{\alpha}$ is taken to be independent of the contact forces and orientations, allowing the $z^{\alpha}$--averaging to be performed and the the force balance equation (\ref{e:force_balance}) rewritten as

\begin{equation}
\delta f^{\rm ext}_{j}
\left\{
\delta_{ij}
-
z
\left\langle
S^{\alpha\beta}_{ij}
\right\rangle_{\alpha\beta}
\right\}
=
0
\label{e:forcebal_sep_z}
\end{equation}

\noindent{}where $z$ is the mean coordination number, and the averaging is now over ${\bf n}^{\alpha\beta}$ and $r^{\alpha\beta}$. Since $\delta{\bf f}^{\rm ext}$ is arbitrary, the quantity inside the brackets in (\ref{e:forcebal_sep_z}) must vanish.

We further assume that the bond orientations are independent of the contact forces. The $\hat{\bf n}^{\alpha\beta}$ are taken to be independent, identically distributed random variables, uniformly distributed over the $d-1$ dimensional unit hypersphere. This neglects any correlations in the topology of the network. It also ignores excluded volume effects, as it allows the same particle to have bonds arbitrarily close together (so the contacting particles would significantly overlap); this should not be crucial near the transition, which is determined by stability requirements rather than excluded volume, but will be relevant at higher~$z$. Performing the average gives

\begin{eqnarray}
z
\Bigg\langle
\frac{d-1}{d}
\left[
1-
\left(
\frac{\lambda f}{r}
\right)^{-1}
\right]^{-1}
&
+
&
\frac{1}{d}
\left[
1-(\lambda f^{\prime})^{-1}
\right]^{-1}
\Bigg\rangle
\nonumber\\
&=&
1
\label{e:forcebal_z_n}
\end{eqnarray}

\noindent{}where the identity $\langle \hat{n}_{i}\hat{n}_{j}\rangle=\frac{1}{d}\delta_{ij}$ has been used.

It remains to specify how the contact forces are distributed. In principle only configurations consistent with force balance should be allowed, but this complication becomes redundant given the approximations leading to~(\ref{e:Sij}), which allows the response from each contact to be calculated independently. A natural choice is then to assume that each contact force $f(r)$, or equivalently each interparticle separation~$r$, are identically and independently distributed according to some given distribution. Clearly this neglects any correlations in the initial force network, but force balance is ensured on average by virtue of the uniform distribution of contact angles already employed. Some calculations for general force distributions will be described later. For now, the simplest choice possible is made, namely a delta--function distribution corresponding to a monodisperse separation $r$, force $f(r)$ and gradient $f^{\prime}(r)\neq0$ for each contact. It is then possible to insert (\ref{e:Sij}) into (\ref{e:forcebal_z_n}) and integrate; the result is finally

\begin{equation}
d\left(\frac{1}{z}-1\right)
=
(d-1)
\frac{1}{\frac{\displaystyle\lambda f(r)}{\displaystyle r} - 1}
-
\frac{1}{1+\lambda\mid f^{\prime}\mid}
\label{e:mma_lambda}
\end{equation}

\noindent{}This is a quadratic equation in~$\lambda$, which is easily solved.

Although $\lambda$ is in principle a measurable quantity, it is more useful to specify results in terms of the shear modulus $G$ or bulk modulus $K$. These can be related to the compliance $\lambda$ by specifying some suitable closed surface $S$ and applying an external force to each particle it cuts, {\em i.e.} forces parallel to $S$ for $G$ and normal for~$K$. Given the displacement of each particle is $\propto\lambda^{-1}$ according to the MMA, it is straightforward to see that

\begin{equation}
G\sim B_{\rm G}\lambda^{-1},
\quad
K\sim B_{\rm K}\lambda^{-1},
\label{e:KGDef}
\end{equation}

\noindent{}where the prefactors $B$ have dimension (length)$^{2-d}$. The relevant length scales are the characteristic length of the enclosing surface $L\sim\sqrt[d-1]{S}$ and the particle radius $r_{0}/2$, but the local closure of the MMA equations means we are unable to determine their weighting in $B_{\rm G}$ and~$B_{\rm K}$. However, if $\lambda$ diverges with~$L$, as is trivially true for a $d=1$ system with fixed boundaries, then $B\sim Lr_{0}^{1-d}$ to ensure finite moduli for arbitrarily large systems. Some form of divergence of $\lambda$ with $L$ is also expected from linear continuum elasticity~\cite{LandauLifshitz}. This subtlety is sidestepped below, where $L$ is assumed to be fixed and finite.

\begin{figure}
\centering
\includegraphics[width=\columnwidth]{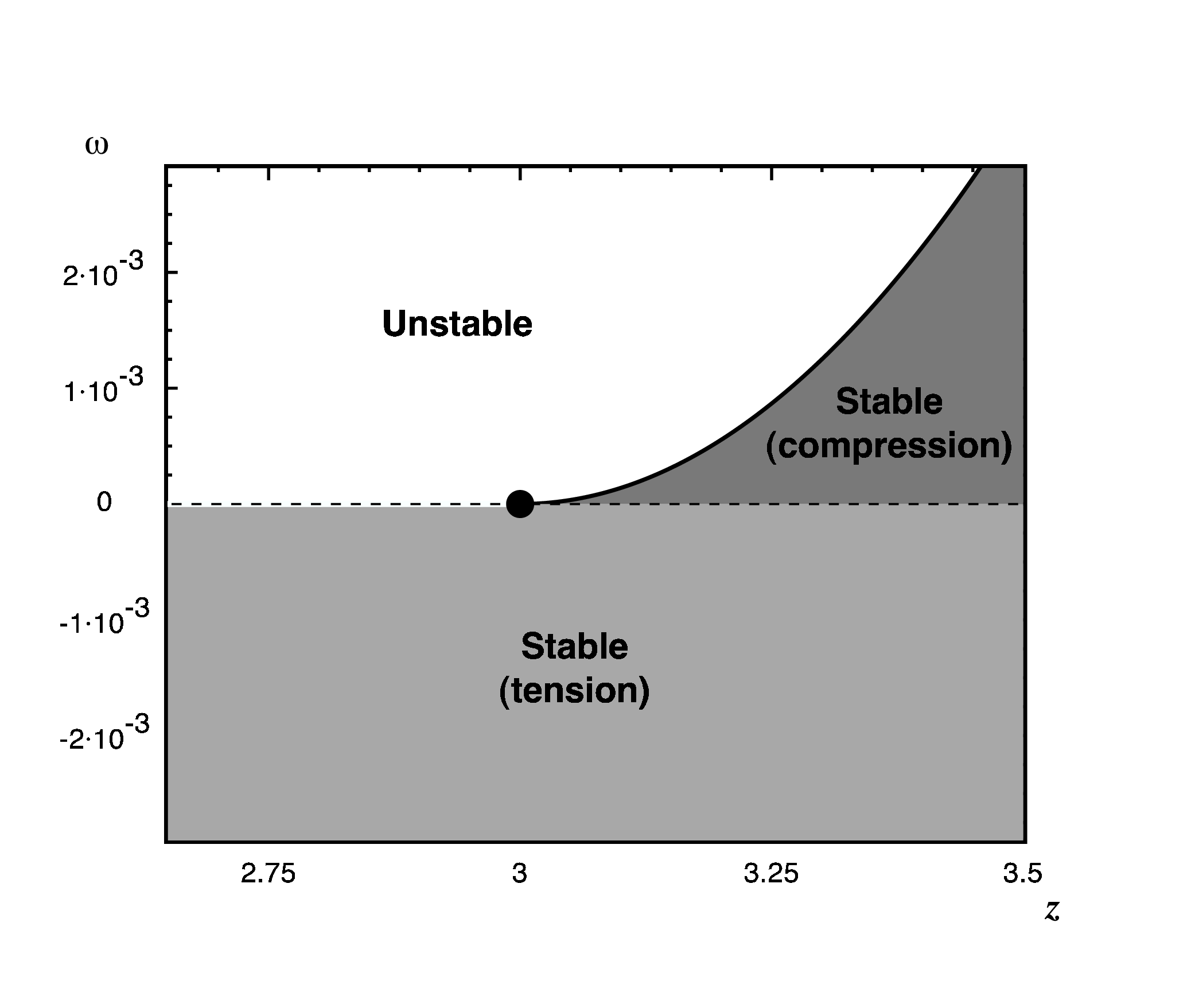}
\caption{Stable regimes of $(z,\omega)$ space in the MMA model, for $d=3$. The black disc at $(d,0)$ is the (unstressed) rigidity percolation transition. All points $z<d$, $\omega=0$ are unstable. For $\omega<0$, corresponding to tensile bonds, the system is always stable (light grey region). For $\omega>0$, corresponding to compression, only the dark grey shaded region is stable.}
\label{f:CompTens}
\end{figure}

\subsection{Stability regimes}

Ignoring the trivial $d=1$ case, dimensionality only enters {\em via} the prefactors and so all $d\geq2$ will be discussed together. The solutions to (\ref{e:mma_lambda}) can be conveniently expressed in terms of the two scalars $z$ and $\omega$, where

\begin{eqnarray}
\omega
&=&
\frac{f(r)/r}{\mid f^{\prime}(r)\mid}
\\
&\sim&
\frac{1}{\fexp}
\left(
1-\frac{r}{r_{0}}
\right)
\label{e:omega}
\end{eqnarray}

\noindent{}is a dimensionless measure of the prestress in the system. The second form (\ref{e:omega}) holds for the particulate potentials (\ref{e:f_r}) in the limit $r\rightarrow r_{0}^{-}$, {\em i.e.} close to the rigidity transition, which is the regime of interest here. A schematic description of the predictions of the MMA are given in Fig.~\ref{f:CompTens}. Some important features are now discussed.

\vspace{\baselineskip}

\noindent{\em $\omega\equiv0$:}
This is the unstressed case in which all contact forces are initially zero, nullifying use of the particulate potentials~(\ref{e:f_r}), which have no linear response at $r=r_{0}$, but still attainable for non--truncated Hookean springs. The equation for $\lambda$ gives the single solution $\lambda|f^{\prime}|=(z/d-1)^{-1}$, or $G,K\sim(z/z_{\rm c}-1)^{f}$ with the transition point $z_{\rm c}=d$ and an exponent $f=1$. The effective medium theory for diluted spring lattices also predicts $f=1$, but at the higher transition point $z_{\rm c}=2d$ in accord with the Maxwell counting estimate~\cite{Feng:85}. The transition value found here, $z_{\rm c}=d$, seems anomalous until one recalls that the basic assumptions of the MMA restrict the motion of the particles to mean forms, thereby reducing their degrees of freedom and hence lowering~$z_{\rm c}$. Despite this, the MMA still predicts a finite transition, and can therefore can be used qualitatively. Any unease over the actual value could be lessened by referring to it as an effective coordination number~$z^{\rm eff}$ if desired.

\vspace{\baselineskip}

\noindent{\em $\omega<0$:}
When all of the bonds are tensile, there is always one real, positive solution of $\lambda$ extending from arbitrarily large $z$ down to a lower value $z_{\rm min}=1$. Again this value is too small; $z_{\rm min}=2$ is more probable, {\em i.e.} infinite chains of particles spanning the system. As $\omega\rightarrow0^{-}$ with $z>z_{\rm c}$ fixed, the single root of $\lambda$ continuously approaches the unstressed solution given above. Repeating this procedure for $z<z_{\rm c}$, however, reveals that $\lambda$ diverges as $|\omega|^{-1}$ and hence $G,K\sim|\omega|$ vanishes continuously as the unstressed axis is reached. Thus just below the $\omega=0$ line, the elastic moduli are very small and the system is inherently weak, becoming weaker as $z$ decreases. This may explain why the few attempts to survey this region in disordered lattices~\cite{Tang:88,Zhou:03} have observed a rapid but gradual crossover of the transition from $z_{\rm c}$ to $z_{\rm min}$: numerical noise and/or arithmetic precision may incorrectly attribute zero values to small but finite moduli.

\vspace{\baselineskip}

\noindent{\em $\omega>0$:}
For compressed bonds, the $(z,\omega)$ plane is partitioned into a stable region with two distinct real, positive roots, and an unstable region for which both roots are either complex or negative. The boundary between the stable and unstable regions is quadratic near $z_{\rm c}$,

\begin{equation}
\omega_{\rm bdy}
\approx
\frac
{(z-z_{\rm c})^{2}}
{4d^{2}(d-1)},
\quad\quad z>z_{\rm c}.
\label{e:zw_boundary}
\end{equation}

\noindent{}Both roots of $\lambda$ coincide on the boundary,

\begin{equation}
\lambda_{\rm bdy}
\approx
\frac{z-z_{\rm c}}{2d(z-1)}
\frac{r_{0}}{\mu}
\left(
\frac{4d^{2}(d-1)}{\fexp(z-z_{\rm c})^{2}}
\right)^{\fexp}
\label{e:lambda_boundary}
\end{equation}

\noindent{}and hence $G_{\rm bdy},K_{\rm bdy}\sim\lambda_{\rm bdy}^{-1}\sim(z-z_{\rm c})^{2\fexp-1}$. Starting from the stable regime and decreasing $\omega$ to zero, one of the roots diverges as $\omega^{-1}$ while the other continuously approaches the unstressed solution, crossing over to become the single root in the tensile regime (where the other root becomes negative).

The manner in which the compressive system becomes unstable is noteworthy. On the boundary, $r_{0}-r\sim(z-z_{\rm c})^{2}$, $f\sim(z-z_{\rm c})^{2\fexp}$ and $f^{\prime}\sim(z-z_{\rm c})^{2(\fexp-1)}$, which according the (\ref{e:Sij}) means that the force transfer is predominantly longitudinal. As already noted by Alexander~\cite{AlexanderRev}, in such cases the change in energy will be positive, from which he infers the system should be stable. However, there are other ways of buckling. An established alternative is a bifurcation to a different class of solution~\cite{Libai&Simmonds}; we might also speculate that the energy landscape may exhibit discontinuities in the limit of infinite system size, allowing some form of catastrophic buckling. In fact, the buckling as envisaged by Alexander, which corresponds to $\lambda<0$ here, {\em does} arise within the MMA, but only for $z<z_{\rm c}$ and small $\omega>0$. The upper boundary in Fig.~\ref{f:CompTens} rather corresponds to when $\lambda$ becomes complex.


\section{Dynamics: Energy minimisation}
\label{s:min}

A system not in a shaded region in Fig.~\ref{f:CompTens} will destabilise under any non--zero noise, evolving its contact network according to the dynamical particle interactions and hence allowing $z$ and~$\omega$ to vary. It can only come to rest in a mechanically stable region. Indeed for sufficiently damped interactions, as assumed here, kinetic arrest can be identified with the point at which the system first touches a stable region. Strong damping also means that the kinetic energy is always small, so that the system will evolve to minimise some suitably defined energy potential. For constant volume $V$, this potential is the internal energy $U$, here just the total potential energy stored in the interparticle bonds. For controlled pressure, the corresponding potential is the enthalpy $H=U+PV$~\cite{Weiner:Book}.

A crucial problem in integrating the fixed $P$ or $V$ dynamics and the $(z,\omega)$ stability diagram is writing down expressions relating $V$ to $z$ and~$\omega$. This is likely to be a subtle issue; under gentle shaking, the particles may form spatially extended structures that would necessitate many--particle variables to calculate~$V$~\cite{Edwards:05,Barker:93,Head:00}, which is clearly beyond the one--particle closure of the MMA equations. For now we ignore such potential pitfalls, and instead assume the following, one--particle description, in the expectation that it will hold in the initial dynamic phase from a highly excited initial state. Simply assume that $V$ is a decreasing function of both $z$ and~$\omega$, as might be expected for uniform, global changes of these variables. The precise choice of $V(z,\omega)$ should incorporate the large changes in $z$ that are possible for small changes in $r$ when the particles are barely touching, {\em i.e.} when $\omega\ll1$. This can be written as

\begin{equation}
\frac{V,_{z}}{V,_{\omega}}
=
D\omega^{b}
\quad
{\rm as}
\quad
\omega\rightarrow 0
\label{e:exponent_b}
\end{equation}

\noindent{}where the unknown exponent $b$ is assumed here to obey $b\geq1$ ($b<1$ alters the scaling behaviour of $V$ with $z$ described later, but not of $P$, $G$ or $K$). The dimensionless constant $D>0$ is some material--dependent parameter.

It is straightforward to derive the internal energy $U$ by employing the same approximations as used to perform the integration of the MMA equations, namely a constant $\omega=\frac{1}{\fexp}(1-r/r_{0})$ and isotropic bond orientations $\hat{\bf n}$,

\begin{equation}
U
=
\frac{Nz}{2}
\frac{r_{0}\mu}{\fexp+1}
(\fexp\omega)^{\fexp+1}
\label{e:U}
\end{equation}

\noindent{}which is the total number of contacts $Nz/2$ multiplied by the bond potential. Similarly, the isotropic pressure $P\delta_{ij}$ is the sum of the $r_{i}f_{j}$ for each bond, divided by the volume~$V$, or (for $r\approx r_{0}$)

\begin{equation}
PV
=
\frac{Nz}{2}
\frac{r_{0}\mu}{d}
(\fexp\omega)^{\fexp}
\label{e:PV}
\end{equation}

\noindent{}where the identity $\langle {\hat n}_{i}{\hat n}_{j}\rangle=\frac{1}{d}\delta_{ij}$ has been used.

Performing the minimisation for both cases reveals broadly the same behaviour; the system will evolve in the direction of increasing $z$ and decreasing $\omega$, as schematised in Fig.~\ref{f:Min_PV} and already discussed in the introduction. For example, for fixed~$V$, the extremum of $U$ (which we assume is the minimum) is found by solving $dU=0$ simultaneously with $dV=0$, where the latter gives the constraint of constant volume. This can be rearranged to give $U,_{\omega}V,_{z}=U,_{z}V,_{\omega}$ and hence from (\ref{e:U}),

\begin{equation}
z\omega^{a}V,_{z}
=
\frac{\omega^{\fexp+1}}{\fexp+1}
V,_{\omega}
\label{e:min_constV}
\end{equation}

\noindent{}This admits the single solution $\omega=0$ in the small--$\omega$ regime of interest here, so $U$ is minimised when all particles are at the limits of their interaction potentials (or at their natural lengths for Hookean springs). Given that the system is constrained to move on lines of constant~$V$~(\ref{e:exponent_b}), the minimum corresponds to a divergent~$z$, clearly unobtainable in a real system. Excluded volume and ordering effects must be incorporated into the theory before any large--$z$ treatment can be attempted. Repeating the enthalpy minimisation at fixed $P$ gives essentially the same behaviour.


\subsection{Kinetic arrest and scaling behaviour}

\begin{figure}
\centering
\includegraphics[width=\columnwidth]{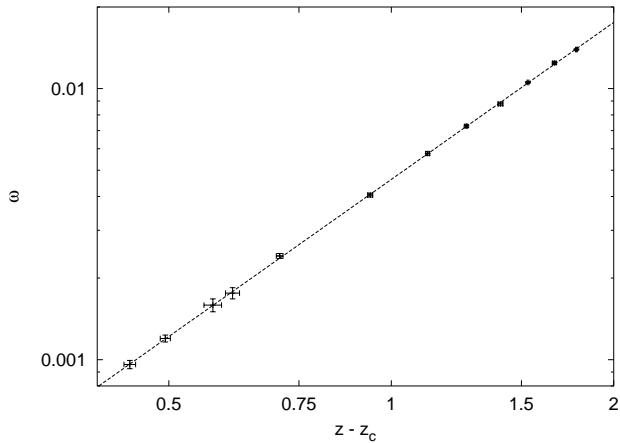}
\caption{Confirmation of the predicted scaling $\langle\omega\rangle\sim(z-z_{\rm c})^{2}$ for $d=3$ Hertzian spheres relaxed at constant volume. The straight line is a fit to $\omega\propto(z-z_{\rm c})^{h}$ with $h=1.92(6)$ and $z_{\rm c}=5.86(4)$, where figures in brackets denote single standard errors on the last digit and $z_{\rm c}<2d=6$ since (rigidly) disconnected rattler particles were included in the average. The simulations were performed on an $N=1000$ monodisperse particle system using the conjugate gradient minimisation routine in the COGNAC module of the open source numerical suite OCTA~\cite{OCTA}.
}
\label{f:Z_Omega}
\end{figure}

Given minimisation drives the system in the direction of small $\omega$ and large $z$, they will enter the mechanically stable region of the $(z,\omega)$ diagram Fig.~\ref{f:Min_PV} somewhere along the stability boundary. For overdamped dynamics, we assume it also stops there, allowing the scaling behaviour of various quantities with $z-z_{\rm c}$ to be determined. According to (\ref{e:zw_boundary}), the microscopic variable $\omega\sim(z-z_{\rm c})^{2}$, which is confirmed by $d=3$ numerical simulation of Hertzian spheres shown in Fig.~\ref{f:Z_Omega}. The relation between the elastic modulus $G$ and $z-z_{\rm c}$ on the boundary has already been given~(\ref{e:lambda_boundary}),

\begin{equation}
G
\sim
(z-z_{\rm c})^{\Kzexp}
\label{e:Kzexp}
\end{equation}

\noindent{}with $\Kzexp=2\fexp-1$. Note that this diverges as $\fexp\rightarrow\frac{1}{2}^{+}$, signifying the breakdown of linear response in this admittedly atypical class of pair potentials.
At the transition point $(z,\omega)=(z_{\rm c},0)$, the pressure $P$ is zero with a finite volume $V_{0}$, so $PV\approx PV_{0}$ to first order and scaling relations involving $P$ can be found independently of the choice of~$V$~(\ref{e:exponent_b}). From (\ref{e:PV}),

\begin{equation}
P
\sim
(z-z_{\rm c})^{\Pzexp}
\label{e:Pzexp}
\end{equation}

\noindent{}with $\Pzexp=2\fexp$. Finally, for relations involving $V$, and hence the volume fraction $\phi\sim V^{-1}$, we find

\begin{equation}
V_{0}-V
\sim
(z-z_{\rm c})^{\Vzexp}
\end{equation}

\noindent{}with $\Vzexp=2$ when $b\geq1$ ($\Vzexp=2b$ for $b<1$). These results are summarised in Table~\ref{t:exponents}, where they are compared to simulation results on various central force systems.

It should be noted that the postulated forms for $V_{0}-V$ and $P$ already allow the bulk modulus~$K$ to be determined without recourse to calculating the mechanical response, and as expected (given the one--bond expressions for $P$ and $V$), the predicted exponent is that of an affine deformation, $K\sim P/(V_{0}-V)\sim(z-z_{\rm c})^{2\alpha-2}$, different to that of $G$ given above. These different exponents have also been seen in simulations~\cite{OHern:03}, although it counters the consensus of lattice models, where all elastic moduli obey the same scaling behaviour~\cite{Head:03,Bergman:84,Feng:84,Arbabi:88}. It also disagrees with the MMA exponents presented earlier~(\ref{e:KGDef}). It is not yet clear if this is a real physical phenomenon or a coincidence of anomalous simulation results and naive choices for $P$ and $V$ in this theory. In any case, we are forced to conclude that the MMA predictions only apply to non--affine deformations.

The similarity of the MMA exponent for $G$ and that of simulations is noteworthy, as the theory relied on a local closure of equations and hence cannot incorporate the long wavelength modes observed in simulations~\cite{Silbert:05}. Persisting with the continuous phase transition analogy, it is possible that the scaling regime for non--trivial exponents is incredibly narrow; this is not without precedent, as ionic fluids exhibit mean field exponents except very near the critical temperature~\cite{Fisher:96,Luijten:02}. Alternatively the upper critical dimension may simply be~2. It should be noted that most (but not all~\cite{OHern:03}) simulations maintain a fixed system size, which will also give mean field exponents when the correlation length exceeds the system size.

A brief survey of other simulations not shown in the table indicate what new features will alter the scaling behaviour and hence represent relevant perturbations. Friction is an obvious candidate, and indeed recent simulations of Zhang {\em et al.}~\cite{Zhang:05} demonstrate that infinite friction will alter the exponents. They also show that finite friction introduces history dependency, suggesting a more advanced description of the dynamical phase will be required for a full theory. Also, the molecular dynamics simulations of Kasahara and Nakanishi~\cite{Kasahara:04a,Kasahara:04b} appear to find exponents consistent with simple rationals that are nonetheless different to those predicted here. This may be because their system has gravity, and hence the contact network is anisotropic and the force balance equations couple directly to an external field, either of which may be a relevant perturbation.

\subsection{Distributed contact forces}

Of all the enhancements to the MMA theory that could be incorporated, perhaps the most pressing is to relax the assumption that every contact force~$f$, or equivalently every particle overlap $\delta=r_{0}-r$, is the same. Simulations have demonstrated that these distributions are in fact continuously distributed right down to zero forces or overlaps~\cite{Silbert:02,Ouaguenouni:97,OHern:01}. Below we present some calculations that probe the effect of polydisperse~$\delta$ within the MMA framework. Our conclusion is that it in fact makes very little difference to the overall behaviour of the system, and does not change the exponents already quoted. We are also able to confirm the scaling of the overlap distribution as observed in simulations~\cite{OHern:03}.

The simplest way to incorporate distributed overlaps is to retain the approximations leading to (\ref{e:forcebal_z_n}), namely independent isotropic bond orientations $\hat{\bf n}$ and a local coordination number independent of the contact forces, and perform the integral over a known, fixed distribution~$P(\delta)$. This can be performed explicitly for a convenient choice of parameters, for instance Hookean interactions $\fexp=1$ with a uniform overlap distribution  $P(\delta)=1/\delta_{0}$ for $0\leq\delta\leq\delta_{0}$. This then produces an equation for $\lambda$ that collapses into the form already studied (\ref{e:mma_lambda}) in the limit  $\delta_{0}\rightarrow0$, with the separation $r$ replaced by a mean overlap $r_{0}-\delta_{0}/2$. Given the volume function (\ref{e:exponent_b}) still holds, we see that $\delta$ being distributed uniformly down to $\delta=0$ does not significantly alter the statics or dynamics of the system in this case, but merely modifies the prefactors.

More general distributions can be considered by assuming that the behaviour observed for the monodisperse case still broadly applies. Specifically, we assume that a unique stressless rigidity transition exists at some point $z=z_{\rm c}$, and a boundary between stable and unstable compressive regimes extends continuously from this point into the region $z>z_{\rm c}$, similar to Fig.~\ref{f:CompTens}. Furthermore, we assume that the system becomes kinetically arrested on this boundary under energy minimisation. Therefore the averaged compliance $\lambda$ is expected to scale with the distance from the transition $\varepsilon=(z-z_{\rm c})/z_{\rm c}>0$ as

\begin{equation}
\lambda
=
\lambda_{0}
\varepsilon^{-\polyexptwo}
\label{e:lambda_ansatz}
\end{equation}

\noindent{}for small $\varepsilon$, with $\polyexptwo$ an unknown positive exponent. Then we make the following scaling {\em ansatz} for the distribution of overlaps $P(\delta)$,

\begin{equation}
P(\delta)
=
\varepsilon^{-\polyexp}
q(\varepsilon^{-\polyexp}\delta)
\label{e:polyexp}
\end{equation}

\noindent{}where $q(x)$ is a fixed distribution. (\ref{e:polyexp}) states that the distribution of overlaps will uniformly contract as the transition is approached, with a width $s$ that vanishes as $s\sim\varepsilon^{\gamma}$ with $\gamma>0$.

Inserting (\ref{e:lambda_ansatz}) and (\ref{e:polyexp}) into (\ref{e:forcebal_z_n}) allows the integration to be performed. Different results occur for different combinations of exponents, but only the form of solution for $\fexp\polyexp>\polyexptwo$ and $\polyexp(\fexp-1)<\polyexptwo$ admits a stressless rigidity transition. The equation for $\lambda_{0}$ in this case is

\begin{eqnarray}
1-\frac{d}{z}
&=&
(d-1)
\frac{\lambda_{0}\mu}{r_{0}}
\varepsilon^{-\polyexptwo+\fexp\polyexp}
\left\langle x^{\fexp}\right\rangle_{q(x)}
\nonumber\\
&+&
\frac{1}{\lambda_{0}\mu\fexp}
\varepsilon^{\polyexptwo-\polyexp(\fexp-1)}
\left\langle
x^{1-\fexp}
\right\rangle_{q(x)}
\label{e:poly_lambda}
\end{eqnarray}

\noindent{}where the angled brackets here denote averaging over the fixed distribution~$q(x)$. (\ref{e:poly_lambda}) admits a solution $\varepsilon=0$ at $z=z_{\rm c}=d$ given the exponent inequalities just quoted. It is similar to the monodisperse expression (\ref{e:mma_lambda}) with $f$ replaced by $\left\langle x^{\fexp}\right\rangle_{q(x)}$, and $\left\langle x^{1-\fexp}\right\rangle_{q(x)}$ the quantity related to $f^{\prime}$.

Stable solutions of (\ref{e:poly_lambda}) correspond to $\lambda_{0}$ real and positive, and so the boundary between the stable and unstable regions can be found using the familiar quadratic equation formulae. For consistency with the earlier assumption that the compressed stability boundary for $\left\langle x^{\fexp}\right\rangle_{q(x)}>0$
is continuously connected to the point $z=z_{\rm c}$, $\varepsilon=0$, we find it is necessary to impose $\gamma=2$. This means that the width of the overlap distribution scales as $s\sim(z-z_{\rm c})^{2}$ near the transition. The simulations of O'Hern {\em et al.} have shown that $s\sim(\phi-\phi_{\rm c})^{\Delta}$ with $\Delta$ close to 1 for Hookean interactions $\fexp=1$ in $d=3$~\cite{OHern:03}. According to Table~\ref{t:exponents}, this corresponds to $s\sim(z-z_{\rm c})^{2\Delta}$, in agreement with the prediction $\gamma=2\Delta=2$ found here. A second consistency check is that $\lambda_{0}={\mathcal O}(1)$, which allows the second exponent to be fixed, $\nu=2\fexp-1$. Note that both of these exponents are equal to their counterparts in the monodisperse--$\delta$ case, {\em i.e.} (\ref{e:zw_boundary}) for $\gamma$ and (\ref{e:lambda_boundary}) for $\nu$. Although this analysis in non--rigourous, it strongly suggests that polydisperse contacts does not alter the scaling picture presented earlier.

 \begin{table}
\caption{\label{t:exponents}Table of the scaling relations between $\Delta z=z-z_{\rm c}$, $G$, $P$ and $\Delta V=V_{0}-V\sim(\phi-\phi_{0})$ (with $\phi_{0}$ the critical volume fraction) as predicted by the MMA theory, where $\alpha$ is the force law exponent~(\ref{e:f_r}), $d$ dimension. For comparison, results from simulations of central force systems and the trivial (affine) predictions are also shown.}
\begin{ruledtabular}
\begin{tabular}{c@{\quad}|@{\quad}ccc}
Model &
$G\sim\Delta z^{\Kzexp}$ &
$P\sim\Delta z^{\Pzexp}$ &
$\Delta V\sim\Delta z^{\Vzexp}$
\\
\hline
MMA\\
$\alpha>0$, $d\geq2$ & $2\alpha-1$ & $2\alpha$ & 2 \\
\hline
Affine (see~\cite{OHern:03}) & $2\alpha-2$ & $2\alpha$ & -\\ 
\hline
Wet foam~\cite{Durian:Foam} \\
$\alpha=1$, $d=2$ & $\approx1$\footnote{Result for shear modulus shown.}  & $2\pm0.4$ & $2\pm0.4$ \\
\hline
O'Hern {\em et al.}~\cite{OHern:03} \\
$\alpha=1$, $d=2,3$ & $1.01\pm0.1^{a}$ & $2.1\pm0.2$ & $2.04\pm0.1$ \\
$\alpha=3/2$, $d=2,3$ & $2.08\pm0.1^{a}$ & $3.15\pm0.3$ & $2.08\pm0.1$ \\
\hline
Zhang {\em et al.}~\cite{Zhang:05} \\
$\alpha=1.28$, $d=3$ & - & $\approx2.45$ & $\approx1.96$ \\
\hline
Makse {\em et al.}~\cite{Makse:05}\footnote{Only frictionless data shown.} \\
$\alpha=3/2$, $d=3$ & - & $3.3\pm0.5$ & $2.1\pm0.6$ \\
\end{tabular}
\end{ruledtabular}
\end{table}

\section{Discussion}
\label{s:discussion}

A central feature of this class of problem is the intrinsic interweaving between the dissipative dynamics as the system cools, with the mechanical response of the arrested state. In this paper, a minimal coupling has been presented, namely that the dynamics proceeds in an independent--particle manner until a mechanically stable region is reached, Fig.~\ref{f:Min_PV}. This is highly simplified, and a more elaborate theory is desirable, perhaps along the lines of the bootstrap percolation model approach~\cite{Schwarz:05}. Intuitively, we expect that, during the dynamic phase, transient overconstrained, rigid clusters will become stressed by interparticle collisions, and relieve this stress by becoming non--rigid, {\em i.e.} expanding into neighbouring, underconstrained regions. Kinetic arrest occurs when a spanning rigid cluster forms. Note that this argument suggests a dynamic homogenisation process, perhaps explaining the apparent appearance of mean field exponents in $d=2$ and 3 simulations in table~\ref{t:exponents}.

This first application of the mean mode approximation has been applied to arguably the simplest particulate problem, namely repulsive central forces in an isotropic system. The simplicity of its results suggests that additional features could be included while remaining tractable. For instance, friction and gravity would be needed before any sensible comparison with real granular media could be made. A problem that may emerge is closing the equations; the averaged force balance equation (\ref{e:force_balance}) only gives one scalar equation, which is why the proposed displacement modes were parameterised by single scalar~$\lambda$. If a future application had too few equations, one possible approach would be to assume the response will minimise the increase in elastic energy, converting it to a minimisation problem with any known equations as constraints. In principle, displacement modes with any number of unknown parameters could be introduced by this approach.

\begin{acknowledgments}

DAH was jointly funded by the European Union Marie Curie program and the JSPS (Japanese Society for the Promotion of Science) program. The author would also like to thank David Dean for suggesting references~\cite{Fisher:96,Luijten:02}.

\end{acknowledgments}

\end{document}